\documentclass[english]{article}
\usepackage{lmodern}

\usepackage[T1]{fontenc}
\usepackage[latin9]{inputenc}
\usepackage{geometry}
\usepackage{babel}
\usepackage{array}
\usepackage{verbatim}
\usepackage{varioref}
\usepackage{multirow}
\usepackage{amsmath}
\usepackage{amssymb}
\PassOptionsToPackage{normalem}{ulem}
\usepackage{ulem}
\usepackage[unicode=true,pdfusetitle,
 bookmarks=true,bookmarksnumbered=true,bookmarksopen=true,bookmarksopenlevel=1,
 breaklinks=true,pdfborder={0 0 0},backref=false,colorlinks=false]
 {hyperref}

\makeatletter

\providecommand{\tabularnewline}{\\}

\newcommand{\code}[1]{\texttt{#1}}

\@ifundefined{date}{}{\date{}}
\makeatother

\begin{document}

\title{\noindent Induction, Coinduction, and Fixed Points\\
in Order Theory, Set Theory, Type Theory, First-Order Logic,\\
and Category Theory: A Concise Comparative Survey}

\author{Moez A. AbdelGawad\\
\code{moez@cs.rice.edu}}
\maketitle
\begin{abstract}
In this survey article (which hitherto is%
\begin{comment}
n't finished but
\end{comment}
{} an ongoing work-in-progress) we present the formulation of the induction
and coinduction principles using the language and conventions of each
of order theory, set theory, programming languages' type theory,
first-order logic, and category theory, for the purpose of examining
some of the similarities and, more significantly, the dissimilarities
between these various mathematical disciplines, and hence shed some
light on the precise relation between these disciplines.

Towards that end, in this article we %
\begin{comment}
get to 
\end{comment}
discuss plenty of related concepts, %
\begin{comment}
including ones 
\end{comment}
such as fixed points, pre-fixed points, post-fixed points, inductive
sets and types, coinductive sets and types, %
\begin{comment}
recursive and corecursive functions (catamorphisms and anamorphisms), 
\end{comment}
algebras and coalgebras. We conclude the survey by hinting at the
possibility of a more abstract and unified treatment that uses concepts
from category theory such as monads and comonads.
\end{abstract}

\section{\label{sec:Introduction}Introduction}

Fixed points and concepts related to them, such as induction and coinduction,
have applications in numerous scientific fields. These mathematical
concepts have applications also in social sciences, and even in common
culture, usually under disguise.\footnote{For more on the intuitions and applications of fixed points and related
notions, the interested reader is invited to check the companion tutorial
article~\cite{AbdelGawad2019a}.}

Fixed points, induction and coinduction have been formulated and studied
in various subfields of mathematics, usually using different vocabulary
in each field. Fields for studying these concepts, formally, include
\emph{order theory}, \emph{set theory}, programming languages (PL)
\emph{type theory}\footnote{That is, the study of data types in computer programming languages.},
\emph{first-order logic}, and \emph{category theory}.

In this article---which started as a brief self-note (now summarized
in~$\mathsection$\ref{sec:Comparison-Summary})---we compare the
various formulations of these concepts by presenting how these concepts,
and many related ones---such as \emph{pre-fixed points}, \emph{post-fixed
points}, \emph{inductive sets/types}, \emph{coinductive sets/types},
\emph{algebras}, and \emph{coalgebras}---are defined in each of these
mathematical subfields. As such this article is structured as follows.

In~$\mathsection$\ref{sec:Order-Theory}~(\nameref{sec:Order-Theory})
we start by presenting how these mathematical concepts are formulated
in a natural and simple manner in order theory. Then, in~$\mathsection$\ref{sec:Set/Class-Theory}~(\nameref{sec:Set/Class-Theory}),
we present the standard formulation of these concepts in set theory.
Since PL type theory builds on set theory, we next present the formulation
of these concepts in the theory of types of \emph{functional} programming
languages (which is largely \emph{structurally-typed}) in~$\mathsection$\ref{sub:(Co)Inductive-Types}~(\nameref{sub:(Co)Inductive-Types}),
then, in~$\mathsection$\ref{sub:Object-Oriented-Type-Theory}~(\nameref{sub:Object-Oriented-Type-Theory}),
we follow that by presenting their formulation in the type theory
of \emph{object-oriented} programming languages (which is largely
\emph{nominally-typed}). Building on intuitions gained from the formulation
of the concepts we presented in~$\mathsection$\ref{sec:Set/Class-Theory},
we then suggest in~$\mathsection$\ref{sec:Logic}~(\nameref{sec:Logic})
a formulation of these concepts in first-order logic. Then we present
the most general formulation of these concepts, in the context of
category theory, in~$\mathsection$\ref{sec:Category-Theory}~(\nameref{sec:Category-Theory}).

In~$\mathsection$\ref{sec:Comparison-Summary}~(\nameref{sec:Comparison-Summary})
we summarize the article by presenting tables that collect the formulations
given in the previous sections%
\begin{comment}
 (and also succinctly express the original intent of this article!)
\end{comment}
. We conclude in~$\mathsection$\ref{sec:More-Abstract}~(\nameref{sec:More-Abstract})
by hinting at the possibility of a more abstract, and more unified
formulation of the concepts we discussed in this article, using concepts
from category theory, namely monads and comonads.

It should be noted that in each particular subfield we discuss in
this article we try to use what seems (to us) to be the most natural
terminology in that field. Hence, in agreement with standard mathematical
practice, the same mathematical concepts may have significantly different
names when presented in different fields in this article. Noting the
significant similarity of the mathematical concepts, via abstracting
away from their names, should be a main goal of readers of this article.

It should be noted also that mathematical-but-non-computer-science
readers---presumably interested mainly in comparing ``pure'' mathematical
subdisciplines\emph{, i.e.}, in comparing formulations in order theory,
set theory, first-order logic, and category theory, but not in PL
type theory---may safely skip reading $\mathsection$\ref{sec:Programming-Languages-Theory}.
Those readers may like to also ignore the rightmost column of Table~\ref{tab:Comparison-1}
and the next-to-leftmost column of Table~\ref{tab:Comparison-2}
in~$\mathsection$\ref{sec:Comparison-Summary}. This %
\begin{comment}
skipped 
\end{comment}
material is mainly of interest to PL theorists only.\footnote{PL theorists may also like to read the more in-depth article~\cite{AbdelGawad2019b}.}

\section{\label{sec:Order-Theory}Order Theory}

Order theory---which includes \emph{lattice theory} as a subfield---is
the branch of mathematics where the concepts \emph{fixed point}, \emph{pre-fixed
point}, \emph{post-fixed point}, \emph{least fixed point}, \emph{greatest
fixed point}, \emph{induction}, and \emph{coinduction} were first
formulated in some generality and the relations between them were
proven~\cite{Tarski1955}. Order theory seems to be the simplest
and most natural setting in which these concepts can be defined and
studied. We thus start this article by presenting the order-theoretic
formulation of these concepts in this section.

\subsubsection{Formulation}

Let $\leq$ (`is less than or equal to') be an \emph{ordering} relation---also
called a \emph{partial} \emph{order}---over a set $\mathbb{O}$ and
let $F:\mathbb{O\rightarrow\mathbb{O}}$ be an \emph{endofunction}
over $\mathbb{O}$ (also called a \emph{self-map} over $\mathbb{O}$,
\emph{i.e.}, a function whose domain and codomain are the same set,
thus mapping a set into itself\footnote{We are focused on unary functions in this article because we are interested
in discussing fixed points and closely-related concepts, to which
multi-arity makes little difference. Note that a binary function $F:\mathbb{O\times O\rightarrow O}$
can be transformed into an equivalent unary one $F':\mathbb{O\rightarrow\left(O\rightarrow O\right)}$
via the technique of ``currying'' (also known in logic as \emph{exportation}).
By iterating currying, it can be applied to multi-ary functions, \emph{i.e.},
functions with any finite arity greater than two. Currying does preserve
monotonicity/variance, and it seems currying is applicable to all
fields of interest to us in this article since, in each field, the
objects of that field---\emph{i.e.}, posets, power sets, types, etc.---and
the ``morphisms/arrows'' between these objects seem to form what
is called (in category theory) a \emph{closed monoidal category}.}).

Given a point $P\in\mathbb{O}$, we call %
\begin{comment}
the 
\end{comment}
point $F\left(P\right)$---the image of $P$ under $F$---the `\emph{$F$-image}'\footnote{In this article, nonstandard names (suggested by the author) are single-quoted
like this `---' when first introduced.} of $P$.

A point $P\in\mathbb{O}$ is called a \emph{pre-fixed point} of $F$
if \uline{its $F$-image is less than or equal to it}, \emph{i.e.},
if 
\[
F\left(P\right)\leq P.
\]
(A pre-fixed point of $F$ is sometimes also called an \emph{$F$-closed
point}, `\emph{$F$-lower bounded point}', or `\emph{$F$-large
point}'; or, when disregarding $F$, an \emph{inductive point} or
\emph{algebraic} \emph{point}.) The greatest element of $\mathbb{O}$,
if it exists (in $\mathbb{O}$), is usually denoted by $\top$, and
it is a pre-fixed point of $F$ for all endofunctions $F$. In fact
$\top$, when it exists, is the \emph{greatest} \emph{pre-fixed} \emph{point}
of $F$ for all $F$.

A point $P\in\mathbb{O}$ is called a \emph{post-fixed point} of $F$
if \uline{it is less than or equal to its $F$-image}, \emph{i.e.},
if 
\[
P\leq F\left(P\right).
\]
(A post-fixed point of $F$ is sometimes also called an \emph{$F$-consistent
point}, \emph{$F$-`(upper)' bounded point}, or `\emph{$F$-small
point}'; or, when disregarding $F$, a\emph{ coinductive point} or
\emph{coalgebraic} \emph{point}.) The least element of $\mathbb{O}$,
if it exists (in $\mathbb{O}$), is usually denoted by $\bot$, and
it is a post-fixed point of $F$ for all endofunctions $F$. In fact
$\bot$, when it exists, is the \emph{least post-fixed point }of $F$
for all $F$.

A point $P\in\mathbb{O}$ is called a \emph{fixed point} (or `\emph{fixed
element}') of $F$ if \uline{it is equal to its $F$-image}, \emph{i.e.},
if 
\[
P=F\left(P\right).
\]
As such, a fixed point of $F$ is simultaneously a pre-fixed point
of $F$ and a post-fixed point of $F$.\medskip{}

Now, if $\leq$ is a \emph{complete lattice} over $\mathbb{O}$ (\emph{i.e.},
if $\leq$ is an ordering relation where meets $\wedge$ and joins
$\vee$ of \emph{all} subsets of $\mathbb{O}$ are guaranteed to exist
in $\mathbb{O}$) and if, in addition, $F$ is a \emph{monotonic }endofunction
over $\mathbb{O}$, \emph{i.e.}, if 
\[
\forall X,Y\in\mathbb{O}.X\leq Y\implies F\left(X\right)\le F\left(Y\right),
\]
then $F$ is called a \emph{generating} \emph{function} (or \emph{generator})
and---as was proven by Tarski~\cite{Tarski1955}---the \emph{least
pre-fixed point }of $F$, called $\mu_{F}$, exists in $\mathbb{O}$
(due to the completeness of $\mathbb{O}$) and $\mu_{F}$ is also
the \emph{least fixed point} (or, \emph{lfp}) of $F$ (due to the
monotonicity of $F$), and the \emph{greatest post-fixed point} of
$F$, called $\nu_{F}$, exists in $\mathbb{O}$ (again, due to the
completeness of $\mathbb{O}$) and $\nu_{F}$ is also the \emph{greatest
fixed point} (or, \emph{gfp}) of $F$ (again, due to the monotonicity
of $F$).\footnote{See Table~\vref{tab:Comparison-1} for the definitions of $\mu_{F}$
and $\nu_{F}$ in order theory.} Further, given that $\mu_{F}$ is the \emph{least} pre-fixed point\emph{
}of $F$ and $\nu_{F}$ is the \emph{greatest} post-fixed point\emph{
}of $F$, for any element $P\in\mathbb{O}$ we have:
\begin{itemize}
\item (\emph{induction})\phantom{co} if $F\left(P\right)\leq P$, then
$\mu_{F}\leq P$,\medskip{}
\\
which, in words, means that if $P$ is a pre-fixed/inductive point
of $F$ (\emph{i.e.}, if the $F$-image of $P$ is less than or equal
to it), then point $\mu_{F}$ is less than or equal to $P$, and,
\item (\emph{coinduction}) if $P\leq F\left(P\right)$, then $P\leq\nu_{F}$,\medskip{}
\\
which, in words, means that if $P$ is a post-fixed/coinductive point
of $F$ (\emph{i.e.}, if $P$ is less than or equal to its $F$-image),
then point $P$ is less than or equal to point $\nu_{F}$.
\end{itemize}

\subsubsection{References}

See~\cite{Davey2002,Rom2008}.

\section{\label{sec:Set/Class-Theory}Set Theory}

In set theory%
\begin{comment}
TODO: \footnote{Even though the set-theoretic formulation may hold in \emph{class
theory} (\emph{i.e.}, the study/theory of collections too big to be
sets, which may include the coinductively-defined non-wellfounded
``sets''), we limit ourselves in this article to presenting a purely
set theoretic formulation only.}
\end{comment}
, the set of subsets of any set---\emph{i.e.}, the powerset of the
set---is \emph{always} a complete lattice under the inclusion ($\subseteq$)
ordering. As such, Tarski's result in lattice theory (see~$\mathsection$\ref{sec:Order-Theory})
was first formulated and proved in the specific context of powersets~\cite{Knaster1928},
which present a simple and very basic example of a context in which
the order-theoretic formulation of induction/coinduction and related
concepts can be applied and demonstrated. Given the unique foundational
significance of set theory in mathematics---unparalleled except, arguably,
by the foundational significance of category theory---the set-theoretic
formulation of the induction and coinduction principles are in fact
\emph{the} standard formulations of these principles. The set-theoretic
formulation also forms the basis for very similar formulations of
these concepts in (structural) type theory (see~$\mathsection$\ref{sub:(Co)Inductive-Types})
and in first-order logic (see~$\mathsection$\ref{sec:Logic}).

\subsubsection{Formulation}

Let $\subseteq$ (`is a subset of') denote the \emph{inclusion}
ordering of set theory and let $\in$ (`is a member of') denote
the \emph{membership} relation. Further, let $\mathbb{U}$ be the
partially-ordered set of all subsets of some fixed set $U$, under
the inclusion ordering (as such $\mathbb{U}=\wp\left(U\right)$, where
$\wp$ is the powerset operation, and is always a complete lattice)
and let $F:\mathbb{U}\rightarrow\mathbb{U}$ be an endofunction over
$\mathbb{U}$.

A set $P\in\mathbb{U}$ (equivalently, $P\subseteq U$) is called
an \emph{$F$-closed set} if \uline{its $F$-image is a subset
of it}, \emph{i.e.}, if 
\[
F\left(P\right)\subseteq P.
\]
(An $F$-closed subset is sometimes also called an \emph{$F$-lower
bounded set} or \emph{$F$-large} \emph{set}.) Set $U$, the largest
set in $\mathbb{U}$, is an $F$-closed set for all endofunctions
$F$---in fact $U$ is the \emph{largest $F$-closed set} for all
$F$.

A set $P\in\mathbb{U}$ is called an \emph{$F$-consistent set} if
\uline{it is a subset of its $F$-image}, \emph{i.e.}, if 
\[
P\subseteq F\left(P\right).
\]
(An $F$-consistent subset is sometimes also called an \emph{$F$-(upper)
bounded set}\footnote{From which comes the name \emph{$F$-bounded polymorphism} in functional
programming. (See~$\mathsection$\ref{sub:(Co)Inductive-Types}.) }, \emph{$F$-correct set}, or \emph{$F$-small} \emph{set}.) The empty
set, $\phi$, the smallest set in $\mathbb{U}$, is an $F$-consistent
set for all endofunctions $F$---in fact $\phi$ is the \emph{smallest
$F$-consistent set} for all $F$.

A set $P\in\mathbb{U}$ is called a \emph{fixed point} (or `\emph{fixed
set}') of $F$ if \uline{it is equal to its $F$-image}, \emph{i.e.},
if 
\[
P=F\left(P\right).
\]
As such, a fixed point of $F$ is simultaneously an $F$-closed set
and an $F$-consistent set.

\medskip{}

Now, given that $\mathbb{U}$ is a complete lattice, if, in addition,
$F$ is a monotonic%
\begin{comment}
 (but not necessarily invertible)
\end{comment}
{} endofunction, \emph{i.e.}, if 
\[
\forall X,Y\in\mathbb{U}(\equiv X,Y\subseteq U).X\subseteq Y\implies F\left(X\right)\subseteq F\left(Y\right)
\]
then $F$ is called a sets-\emph{generating function} (or \emph{generator})
and an inductively-defined subset $\mu_{F}$, the \emph{smallest $F$-closed
set}, exists in $\mathbb{U}$, and $\mu_{F}$ is also the \emph{smallest
fixed point} of $F$, and a coinductively-defined subset $\nu_{F}$,
the \emph{largest $F$-consistent set}, exists in $\mathbb{U}$, and
$\nu_{F}$ is also the \emph{largest fixed point} of $F$.\footnote{See Table~\vref{tab:Comparison-1} for the definitions of $\mu_{F}$
and $\nu_{F}$ in set theory.} Further, for any set $P\in\mathbb{U}$ we have:
\begin{itemize}
\item (\emph{induction})\phantom{co} $F\left(P\right)\subseteq P\implies\mu_{F}\subseteq P$\medskip{}
\\
(\emph{i.e., }$\forall x\in F\left(P\right).x\in P\implies\forall x\in\mu_{F}.x\in P$),\medskip{}
\\
which, in words, means that if (we can prove that) set $P$ is an
$F$-closed set, then (by induction, we get that) set $\mu_{F}$ is
a subset of $P$, and
\item (\emph{coinduction}) $P\subseteq F\left(P\right)\implies P\subseteq\nu_{F}$\medskip{}
\emph{}\\
(\emph{i.e., }$\forall x\in P.x\in F\left(P\right)\implies\forall x\in P.x\in\nu_{F}$),\medskip{}
\\
which, in words, means that if (we can prove that) set $P$ is an
$F$-consistent set, then (by coinduction, we get that) $P$ is a
subset of set $\nu_{F}$.
\end{itemize}

\subsubsection{Induction Instances}
\begin{itemize}
\item An instance of the set-theoretic induction principle presented above
is the standard \emph{mathematical induction} principle. In this well-known
instance, induction is setup as follows: $F$ is the ``successor''
function (of Peano)\footnote{Namely, $F\left(X\right)=\left\{ 0\right\} \cup\left\{ x+1|x\in X\right\} $,
where, \emph{e.g.}, $F\left(\left\{ 0,3,5\right\} \right)=\left\{ 0,1,4,6\right\} $.}, $\mu_{F}$ (the smallest fixed point of the successor function $F$)
is the set of natural numbers $\mathbb{N}$, and $P$ is any inductive
property/set of numbers.

\begin{itemize}
\item For an example of an inductive set, the reader is invited to check~$\mathsection$2
of the tutorial article~\cite{AbdelGawad2019a}.
\end{itemize}
\item Another instance of the induction principle is \emph{lexicographic
induction}, defined on lexicographically linearly-ordered (\emph{i.e.},
``dictionary ordered'') pairs of elements~\cite{TAPL,DM}. In $\mathsection$\ref{sub:(Co)Inductive-Types}
we will see a type-theoretic formulation of the induction principle
that is closely related to the set-theoretic one above. The type-theoretic
formulation is the basis for yet a third instance of the induction
principle---called \emph{structural induction}---that is extensively
used in programming semantics and automated theorem proving (ATP),
including reasoning about and proving properties of (functional) software.
\end{itemize}

\subsubsection{Coinduction Instances and Intuition}
\begin{itemize}
\item %
Even though coinduction is the dual of induction, and thus apparently
very similar to it, practical uses of coinduction are relatively obscure
compared to those of induction. For some applications of coinduction
see, \emph{e.g.},~\cite{Brandt1998,Sangiorgi2012,Kozen2016}.
\item Another less-obvious instance of a coinductive set is the standard
subtyping relation in (nominally-typed) object-oriented programming
languages. See the notes of~$\mathsection$2.2 of~\cite{AbdelGawad2019b}
for more details.
\item For developing intuitions for coinduction, coinductive sets, and coinductive
types, the reader is invited to check~$\mathsection$3 of the tutorial
article~\cite{AbdelGawad2019a}.
\end{itemize}

\subsubsection{Illustrations}

In~$\mathsection$4 of the tutorial article~\cite{AbdelGawad2019a}
we present diagrams that use examples from set theory, number theory,
and real analysis to concretely illustrate inductive sets, coinductive
sets, and the other related concepts we discuss in this survey article.

\subsubsection{References}

See~\cite{TAPL,Barwise1996}.

\section{\label{sec:Programming-Languages-Theory}Programming Languages Theory}

Given the `types as sets' view of types in programming languages,
this section builds on the \emph{set}-theoretic presentation in $\mathsection$\ref{sec:Set/Class-Theory}
to present the formulation of the induction and coinduction principles
in the context of programming languages type theory.

\subsection[FP Type Theory]{\label{sub:(Co)Inductive-Types}Inductive and Coinductive Functional
Data Types}

The formulation of the induction and coinduction principles in the
context of functional programming type theory builds on the concepts
developed in $\mathsection$\ref{sec:Set/Class-Theory}. For the motivations
behind this formulation, and more details, see~$\mathsection$2.1
of~\cite{AbdelGawad2019b}.

\subsubsection{Formulation}

Let $\mathbb{D}$ be the set of \emph{structural} types in functional
programming.

\medskip{}

Let $\subseteq$ (`is a subset/subtype of') denote the \emph{structural
subtyping/inclusion} relation between structural data types, and let
$:$ (`has type/is a member of/has structural property') denote
the \emph{structural typing} relation between structural data values
and structural data types.

Now, if $F:\mathbb{D}\rightarrow\mathbb{D}$ is a polynomial (with
powers) datatype constructor, \emph{i.e.}, if 
\[
\forall S,T\in\mathbb{D}.S\subseteq T\implies F\left(S\right)\subseteq F\left(T\right),
\]
then an inductively-defined type/set $\mu_{F}$, the \emph{smallest
$F$-closed} \emph{set}, exists in $\mathbb{D}$, and $\mu_{F}$ is
also the \emph{smallest fixed point} of $F$, and a coinductively-defined
type/set $\nu_{F}$, the \emph{largest $F$-consistent} \emph{set},
exists in $\mathbb{D}$, and $\nu_{F}$ is also the \emph{largest
fixed point} of $F$.\footnote{See Table~\vref{tab:Comparison-1} for the definitions of $\mu_{F}$
and $\nu_{F}$.}%
\begin{comment}
(noting that $\Pi$\textbackslash{}$\bigcap$ is `arbitrary product'
and $\Sigma$\textbackslash{}$\bigcup$ is `arbitrary disjoint union')
\end{comment}

Further, for any type $P\in\mathbb{D}$ (where $P$, as a structural
type, expresses a structural property of data values) we have:
\begin{itemize}
\item (\emph{structural induction, and recursion})\phantom{co} $F\left(P\right)\subseteq P\implies\mu_{F}\subseteq P$\medskip{}
\\
(\emph{i.e., }$\forall p:F\left(P\right).p:P\implies\forall p:\mu_{F}.p:P$),\medskip{}
\\
which, in words, means that if the (structural) property $P$ is \emph{preserved}
by $F$ (\emph{i.e.}, if $P$ is $F$-closed), then all data values
of the inductive type $\mu_{F}$ have property $P$ (\emph{i.e.},
$\mu_{F}\subseteq P$).\\
Furthermore, borrowing terminology from category theory (see~$\mathsection$\ref{sec:Category-Theory}),
a \emph{recursive} function $f:\mu_{F}\rightarrow P$ that maps data
values of the inductive type $\mu_{F}$ to data values of type $P$
(\emph{i.e.}, having structural property $P$) is the unique \emph{catamorphism}
(also called a \emph{fold}) from $\mu_{F}$ to $P$ (where $\mu_{F}$
is viewed as an initial $F$-algebra and $P$ as an $F$-algebra),
and
\item (\emph{structural coinduction, and corecursion}) $P\subseteq F\left(P\right)\implies P\subseteq\nu_{F}$\medskip{}
\emph{}\\
(\emph{i.e., }$\forall p:P.p:F\left(P\right)\implies\forall p:P.p:\nu_{F}),$\medskip{}
\\
which, in words, means that if the (structural) property $P$ is \emph{reflected}
by $F$ (\emph{i.e.}, if $P$ is $F$-consistent), then all data values
that have property $P$ are data values of the coinductive type $\nu_{F}$
(\emph{i.e.}, $P\subseteq\nu_{F}$).\\
Furthermore, borrowing terminology from category theory, a \emph{corecursive}
function $f:P\rightarrow\nu_{F}$ that maps data values of type $P$
(\emph{i.e.}, having structural property $P$) to data values of the
coinductive type $\nu_{F}$ is the unique \emph{anamorphism} from
$P$ to $\nu_{F}$ (where $P$ is viewed as an $F$-coalgebra and
$\nu_{F}$ as a final $F$-coalgebra).
\end{itemize}

\subsubsection{Notes}
\begin{itemize}
\item To guarantee the existence of $\mu_{F}$ and $\nu_{F}$ in $\mathbb{D}$
for all type constructors $F$, and hence to guarantee the ability
to reason easily---\emph{i.e.}, inductively and coinductively---about
functional programs, the domain $\mathbb{D}$ of %
\begin{comment}
(VIP: values domain and types domain are distinct!)
\end{comment}
types in functional programming is \emph{deliberately} constructed
to be a complete lattice under the inclusion ordering%
\begin{comment}
 (or a directed-complete partial order, abbr. CPO, in which also the
coinductively-defined $\mu_{F}$, but not necessarily $\nu_{F}$,
is guaranteed to exist for every monotonic $F$~\cite[Ch. 8]{Davey2002})
\end{comment}
. This is achieved by limiting the type constructors used in constructing
$\mathbb{D}$ and over $\mathbb{D}$ to \emph{structural} type constructors
only (\emph{i.e.}, to the constructors $+$, $\times$, $\rightarrow$
and their compositions, in addition to basic types such as \code{Unit},
\code{Bool}, \code{Top}, \code{Nat} and \code{Int}). More details
can be found in~$\mathsection$2.1 of~\cite{AbdelGawad2019b}.
\end{itemize}

\subsection[OOP Type Theory]{\label{sub:Object-Oriented-Type-Theory}Object-Oriented Type Theory}

The formulation of the induction and coinduction principles in the
context of object-oriented type theory roughly mimics the formulation
in~$\mathsection$\ref{sub:(Co)Inductive-Types}. For the motivations
behind the exact formulation, and more details, see~$\mathsection$2.2
of~\cite{AbdelGawad2019b}.

\subsubsection{Formulation}

Let $<:$ (`is a subtype of') denote the \emph{nominal subtyping}
relation between nominal data types (\emph{i.e.}, class types), and
let $:$ (`has type') denote the \emph{nominal typing} relation
between nominal data values (\emph{i.e.}, objects) and nominal data
types.

Further, let $\mathbb{T}$ be the set of nominal types in object-oriented
programming, ordered by the nominal subtyping relation, and let $F:\mathbb{T}\rightarrow\mathbb{T}$
be a type constructor over $\mathbb{T}$ (\emph{e.g.}, a generic class).\footnote{Unlike poset $\mathbb{D}$ in $\mathsection$\ref{sub:(Co)Inductive-Types}
(of structural types under the structural subtyping relation), poset
$\mathbb{T}$ (of nominal types under the nominal subtyping relation)
is \emph{not} guaranteed to be a complete lattice.}

A type $P\in\mathbb{T}$ is called an `\emph{$F$-supertype}' if
\uline{its $F$-image is a subtype of it}, \emph{i.e.}, if 
\[
F\left(P\right)<:P,
\]
and $P$ is said to be \emph{preserved} by $F$. (An $F$-supertype
is sometimes also called an \emph{$F$-closed type}, \emph{$F$-lower
bounded type}, or \emph{$F$-large type}). The \emph{root }or \emph{top}
of the subtyping hierarchy, if it exists (in $\mathbb{T}$), is usually
called \code{Object} or \code{All}, and it is an $F$-supertype
for all generic classes $F$. In fact the top type, when it exists,
is the \emph{greatest $F$-supertype} for all $F$.

A type $P\in\mathbb{T}$ is called an `\emph{$F$-subtype}' if \uline{it
is a subtype of its $F$-image}, \emph{i.e.}, if 
\[
P<:F\left(P\right),
\]
and $P$ is said to be \emph{reflected} by $F$. (An $F$-subtype
is sometimes also called an \emph{$F$-consistent type}, \emph{$F$-(upper)
bounded type}\footnote{From which comes the name \emph{$F$-bounded generics} in object-oriented
programming.}, or \emph{$F$-small type}). The \emph{bottom} of the subtyping hierarchy,
if it exists (in $\mathbb{T}$), is usually called \code{Null} or
\code{Nothing}, and it is an $F$-subtype for all generic classes
$F$. In fact the bottom type, when it exists, is the \emph{least
$F$-subtype} for all $F$.

A type $P\in\mathbb{T}$ is called a \emph{fixed point} (or `\emph{fixed
type}') of $F$ if \uline{it is equal to its $F$-image}, \emph{i.e.},
if 
\[
P=F\left(P\right).
\]
As such, a fixed point of $F$ is simultaneously an $F$-supertype
and an $F$-subtype. (Such fixed types/points are rare in OOP practice).

\medskip{}

Now, if $F$ is\emph{ }a \emph{covariant} generic class (\emph{i.e.},
a types-generator), \emph{i.e.}, if 
\[
\forall S,T\in\mathbb{T}.S<:T\implies F\left(S\right)<:F\left(T\right),
\]
and if $\mu_{F}$, the `\emph{least $F$-supertype}' exists in $\mathbb{T}$,
and $\mu_{F}$ is also the \emph{least fixed point} of $F$, and if
$\nu_{F}$, the `\emph{greatest $F$-subtype}', exists in $\mathbb{T}$,
and $\nu_{F}$ is also the \emph{greatest fixed point} of $F$,\footnote{See Table~\vref{tab:Comparison-2} for the definitions of $\mu_{F}$
and $\nu_{F}$ in the (rare) case when $\mathbb{T}$ happens to be
a complete lattice.} then, for any type $P\in\mathbb{T}$ we have:
\begin{itemize}
\item (\emph{induction})\phantom{co} $F\left(P\right)<:P\implies\mu_{F}<:P$\medskip{}
\\
(\emph{i.e., }$\forall p:F\left(P\right).p:P\implies\forall p:\mu_{F}.p:P$),\medskip{}
\\
which, in words, means that if the contract (\emph{i.e.}, behavioral
type) $P$ is \emph{preserved} by $F$ (\emph{i.e.}, $P$ is an $F$-supertype),
then the inductive type $\mu_{F}$ is a subtype of $P$, and
\item (\emph{coinduction}) $P<:F\left(P\right)\implies P<:\nu_{F}$\medskip{}
\emph{}\\
(\emph{i.e., }$\forall p:P.p:F\left(P\right)\implies\forall p:P.p:\nu_{F}$),\medskip{}
\\
which, in words, means that if the contract (\emph{i.e.}, behavioral
type) $P$ is \emph{reflected} by $F$ (\emph{i.e.}, $P$ is an $F$-subtype),
then $P$ is a subtype of the coinductive type $\nu_{F}$.
\end{itemize}

\subsubsection{Notes}
\begin{itemize}
\item More on the differences between structural subtyping, as found in
functional programming type theory, and nominal typing, as found in
object-oriented programming type theory, can be found in~$\mathsection$2.2
of~\cite{AbdelGawad2019b}.
\end{itemize}

\section{\label{sec:Logic}First-Order Logic}

Mathematical logic, particularly \emph{first-order logic} (FOL), is
the foundation of set theory and similar theories. Specifically, via
the \emph{Aussonderungsaxiom} and other axioms in set theory (see~\cite[$\mathsection$2]{Halmos60}),
first-order logic is strongly tied to set theory\footnote{The \emph{Aussonderungsaxiom} (German) is also called the subset/separation
axiom or the axiom of specification/comprehension.}, as well as to \emph{class theory} and \emph{non-wellfounded set
theory} (see~\cite{Barwise1996}).

As such, in correspondence with the set-theoretic concepts and definitions
presented in $\mathsection$\ref{sec:Set/Class-Theory}, one should
expect to find counterparts in first-order logic. Even though seemingly
unpopular in mathematical logic literature, we try to explore these
corresponding concepts in this section. The discussion of these concepts
in logic is also a step that prepares for discussing these concepts
in $\mathsection$\ref{sec:Category-Theory} in the more general setting
of category theory.

\subsubsection{Formulation}

Let $\Rightarrow$ (`implies') denote the \emph{implication} relation
between predicates/logical statements of first-order logic\footnote{Note that, as in earlier sections, we use the long implication symbol
`$\implies$' to denote the implication relation in the \emph{metalogic}
(\emph{i.e.}, the logic used to reason \emph{about} objects of interest
to us, \emph{e.g.}, points, sets, types, or---as is the case in this
section---statements of first-order logic). The reader in this section
should be careful not to confuse the metalogical implication relation
(denoted by the long implication symbol $\implies$) with the implication
relation of first-order logic (used to reason \emph{in }FOL, and denoted
by the short implication symbol $\Rightarrow$).}, and let juxtaposition or $\cdot\left(\cdot\right)$ (`is satisfied
by/applies to') denote the \emph{satisfiability} relation between
predicates and objects/elements. Further, let $\mathbb{S}$ be the
set of statements (\emph{i.e.}, the well-formed formulas) of first-order
logic ordered by implication ($\mathbb{S}$ is thus a complete lattice)
and let $F:\mathbb{S}\rightarrow\mathbb{S}$ be a logical operator
over $\mathbb{S}$.\emph{}\footnote{Such as $\land$ {[}and{]} and $\lor$ {[}or{]} (both of which are
covariant/monotonic logical operators), $\lnot$ {[}not{]} (which
is contravariant/anti-monotonic), or compositions thereof.}

A statement $P\in\mathbb{S}$ is called an `\emph{$F$-weak statement}'
if \uline{its $F$-image implies it}, \emph{i.e.}, if 
\[
F\left(P\right)\Rightarrow P.
\]
Statement $True$ is an $F$-weak statement for all endofunctors $F$---in
fact $True$ is the \emph{weakest $F$-weak statement} for all $F$.

A statement $P\in\mathbb{S}$ is called an `\emph{$F$-strong }statement'
if \uline{it implies its $F$-image}, \emph{i.e.}, if 
\[
P\Rightarrow F\left(P\right).
\]
Statement $False$ is an $F$-strong statement for all endofunctors
$F$---in fact $False$ is the \emph{strongest $F$-strong statement}
for all $F$.

A statement $P\in\mathbb{S}$ is called a \emph{fixed point} (or `\emph{fixed
statement}') of $F$ if \uline{it is equivalent to its $F$-image},
\emph{i.e.}, if 
\[
P\Leftrightarrow F\left(P\right).
\]
As such, a fixed point of $F$ is simultaneously an $F$-weak statement
and an $F$-strong statement.

\medskip{}

Now, if $F$ is a \emph{covariant} logical operator (\emph{i.e.},
a statements-generator), \emph{i.e.}, if 
\[
\forall S,T\in\mathbb{S}.\left(S\Rightarrow T\right)\implies\left(F\left(S\right)\Rightarrow F\left(T\right)\right),
\]
then $\mu_{F}$, the `\emph{strongest $F$-weak statement}', exists
in $\mathbb{P}$, and $\mu_{F}$ is also the `\emph{strongest fixed
point}' of $F$, and $\nu_{F}$, the `\emph{weakest $F$-strong
statement}', exists in $\mathbb{P}$, and $\nu_{F}$ is also the
`\emph{weakest fixed point}' of $F$.\footnote{See Table~\vref{tab:Comparison-2} for the definitions of $\mu_{F}$
and $\nu_{F}$.} Further, for any statement $P\in\mathbb{S}$ we have:
\begin{itemize}
\item (\emph{induction})\phantom{co} if $\forall x.F\left(P\right)\left(x\right)\Rightarrow P\left(x\right)$,
then $\forall x.\mu_{F}\left(x\right)\Rightarrow P\left(x\right)$,\medskip{}
\\
which, in words, means that if $P$ is an $F$-weak statement, then
$\mu_{F}$ implies $P$, and
\item (\emph{coinduction}) if $\forall x.P\left(x\right)\Rightarrow F\left(P\right)\left(x\right)$,
then $\forall x.P\left(x\right)\Rightarrow\nu_{F}\left(x\right)$,\medskip{}
\\
which, in words, means that if $P$ is an $F$-strong statement, then
$P$ implies $\nu_{F}$.
\end{itemize}

\subsubsection{References}

See~\cite{AbdelGawad2018f}(!)%
\begin{comment}
{[}TODO{]}
\end{comment}
, and~\cite{EndertonLogic72,Shoenfield67,Krantz02,Santocanale03,Forster2003}.

\section{\label{sec:Category-Theory}Category Theory}

Category theory seems to present the most general context in which
the notions of induction, coinduction, fixed points, pre-fixed points
(called algebras) and post-fixed points (called coalgebras) can be
studied.

\subsubsection{Formulation}

Let $\rightarrow$ (`is related to'/`arrow') denote that two objects
in a category are related (\emph{i.e.}, denote the `\emph{is-related-to}'
relation, or, more concisely, denote relatedness).\footnote{For mostly historical reasons, an arrow relating two objects in a
category is sometimes also called a \emph{morphism}. We prefer using
\emph{is-related-to} (\emph{i.e.}, has some relationship with) or
\emph{arrow} instead, since these terms seem to be more easily and
intuitively understood, and also because they seem to be more in agreement
with the abstract and general nature of category theory.} Further, let $\mathbb{O}$ be the collection of objects of a category
$\mathbb{O}$ (\emph{i.e.}, the category and the collection of its
objects are homonyms%
\begin{comment}
 have the same name, $\mathbb{O}$
\end{comment}
) and let $F:\mathbb{O\rightarrow\mathbb{O}}$ be\footnote{Note that, similar to the situation for symbols $\implies$ and $\Rightarrow$
that we met in $\mathsection$\ref{sec:Logic}, in this section the
same exact symbol $\rightarrow$ is used to denote two (strongly-related
but slightly different) meanings: the first, that two objects in a
category are related (which is the meaning specific to category theory),
while the second is the functional type of a self-map/endofunction/endofunctor
$F$ that acts on objects of interest to us (\emph{i.e.}, points,
sets, types, etc.), which is the meaning for $\rightarrow$ that we
have been using all along since the beginning of this article.} an endofunctor over $\mathbb{O}$.

An object $P\in\mathbb{O}$ is called an \emph{$F$-algebra} if \uline{its
$F$-image is related to it}, \emph{i.e.}, if 
\[
F\left(P\right)\rightarrow P.
\]
An object $P\in\mathbb{O}$ is called an \emph{$F$-coalgebra} if
\uline{it is related to its $F$-image}, \emph{i.e.}, if 
\[
P\rightarrow F\left(P\right).
\]

\medskip{}

Now, if $F$\emph{ }is a \emph{covariant} endofunctor, \emph{i.e.},
if 
\[
\forall X,Y\in\mathbb{O}.X\rightarrow Y\implies F\left(X\right)\rightarrow F\left(Y\right),
\]
and if an \emph{initial $F$-algebra} $\mu_{F}$ %
\begin{comment}
TODO: Not counterpart of least \emph{fixed} point, but of least \emph{pre-fixed}
point. See discussion on \code{F<?>} in $\mathsection$\ref{sub:Object-Oriented-Type-Theory}
(Footnote~\ref{fn:gfsub-lfsup}). 
\end{comment}
exists in $\mathbb{O}$ and a \emph{final $F$-coalgebra} $\nu_{F}$
\begin{comment}
TODO: Not counterpart of greatest \emph{fixed} point, but of greatest
\emph{post-fixed} point. See discussion on \code{F<?>} in $\mathsection$\ref{sub:Object-Oriented-Type-Theory}.
(Footnote~\ref{fn:gfsub-lfsup}). 
\end{comment}
exists in $\mathbb{O}$,\footnote{\label{fn:ct-gfsub-lfsup}While referring to $\mathsection$\ref{sec:Order-Theory},
note that a category-theoretic initial $F$-algebra and final $F$-coalgebra
are \emph{not} the exact counterparts of an order-theoretic least
fixed point and greatest fixed point but of a least \emph{pre}-fixed
point and greatest \emph{post}-fixed point. This slight difference
is significant, since, for example, a least pre-fixed point is \emph{not}
necessarily a least fixed point, unless the underlying ordering is
a complete lattice (or at least is a meet-complete lattice), and
also a greatest post-fixed point is \emph{not} necessarily a greatest
fixed point, unless the underlying ordering is a complete lattice
(or at least is a join-complete lattice). The difference is demonstrated,
for example, by the subtyping relation in generic nominally-typed
OOP (see Footnote~14 in~$\mathsection$2.2 of~\cite{AbdelGawad2019b}).}%
\begin{comment}
TODO: , or so do things seem to us as of the time of this writing
(Dec. 2018/Jan. 2019). By mid Jan. it now seems to us more likely
that $\mu_{F}$ and $\nu_{F}$ can also be used to denote least \emph{pre}-fixed
points and greatest \emph{post}-fixed points. What if $F$ is not
monotonic?
\end{comment}
{} then for any object $P\in\mathbb{O}$ we have:
\begin{itemize}
\item (\emph{induction})\phantom{co} if $F\left(P\right)\rightarrow P$,
then $\mu_{F}\rightarrow P$,\medskip{}
\\
which, in words, means that if $P$ is an \emph{$F$-algebra}, then
$\mu_{F}$ is related to $P$ (via a unique ``complex-to-simple''
arrow called a \emph{catamorphism}), and
\item (\emph{coinduction}) if $P\rightarrow F\left(P\right)$, then $P\rightarrow\nu_{F}$,\medskip{}
\\
which, in words, means that if $P$ is an \emph{$F$-coalgebra}, then
$P$ is related to $\nu_{F}$ (via a unique ``simple-to-complex''
arrow called an \emph{anamorphism}).
\end{itemize}

\subsubsection{Notes}
\begin{itemize}
\item Even though each of an initial algebra and a final coalgebra is simultaneously
an algebra and a coalgebra, it should be noted that there is no explicit
concept in category theory corresponding to the concept of a fixed
point in order theory, due to the general lack of an equality relation
in category theory and the use of the isomorphism relation instead.

\begin{itemize}
\item If such a ``fixedness'' notion is defined in category theory, it
would denote an object that is simultaneously an algebra and a coalgebra,
\emph{i.e.}, for a functor $F$, a ``fixed object'' $P\in\mathbb{O}$
of $F$ will be related to the object $F\left(P\right)\in\mathbb{O}$
\emph{and} vice versa. This usually means that $P$ and $F\left(P\right)$,
if not the same object, are \emph{isomorphic} objects. (That is in
fact the case for any initial $F$-algebra and for any final $F$-coalgebra,
which---given the uniqueness of arrows from an initial algebra and
to a final coalgebra---are indeed isomorphic to their $F$-images).
\end{itemize}
\item Note also that, unlike the case in order theory, the induction and
coinduction principles can be expressed in category theory using a
``point-free style'' (as we do in this article) but they can be
also expressed using a ``point-wise style''\footnote{By giving a name to a specific arrow that relates two objects of a
category, \emph{e.g.}, using notation such as $f:X\rightarrow Y$
to mean not only that objects $X$ and $Y$ are related but also that
they are related by a particular arrow named $f$.}. As such, regarding the possibility of expressing the two principles
using either a point-wise style or a point-free style, category theory
agrees more with set theory, type theory, and (first order) logic
than it does with order theory.
\item Incidentally, categories are more similar to \emph{preorders} (sometimes,
but not invariably, also called \emph{quasiorders}) than they are
similar to partial-orders (\emph{i.e.}, posets). This is because a
category, when viewed as an ordered set, is not necessarily anti-symmetric.

\begin{itemize}
\item Categories are more general than preorders however, since a category
can have \emph{multiple} arrows between any pair of its objects, whereas
any two elements/points of a preorder can only either have one ``arrow''
between the two points (denoted by $\leq$ rather than $\rightarrow$)
or not have one. (This possible multiplicity is what enables, and
sometimes even necessitates, the use of arrow names, so as to distinguish
between different arrows when multiple arrows relate the same pair
of objects). As such, every preorder is a category, or, more precisely,
every preorder has a unique (small) category---appropriately called
a $(0,1)$-category, a \emph{thin} category, or a \emph{bool}(\emph{ean})-category---corresponding
to it, and vice versa. However, generally-speaking, while there is
a unique preorder corresponding to each category but there is \emph{not}
a unique category corresponding to each preorder (typically many categories
correspond to a single preorder $O$, including the thin category
corresponding to $O$).%
\begin{comment}
TODO: Revise. \footnote{Even further, in fact there exists (as the left adjoint of an adjunction)
a functor, called \emph{embed},\emph{ }that maps a preorder to its
corresponding unique $(0,1)$-category (\emph{i.e.}, it embeds the
category of preorders into the category of categories). The right
adjoint of this functor, called \emph{strip}, maps any category, with
some information loss, to a canonical preorder, \emph{i.e.}, it non-injectively
may map multiple distinct categories to the same preorder (specifically,
\emph{strip }maps two different categories that have isomorphic collections
of objects---which, particularly, do not differ in the count of their
objects---but that may differ only in the non-zero count of arrows
between all pairs of corresponding objects to the same preorder).
Further, there is another functor, called \emph{clique}, that, non-injectively
again, maps a preorder to a canonical poset (by mapping distinct but
equivalent/isomorphic points/elements of a preorder to the same element/point
of a poset). As is immediately clear, any poset can be mapped to (\emph{i.e.},
viewed as) a category via an inclusion/embedding. By composing functors
\emph{clique} after \emph{strip}, any category can also be mapped,
but with some information loss, to a canonical poset. See~\cite{spivak2014category}Section~??
for more details.}
\end{comment}
\footnote{Indeed the relation between the two fields is precise to the extent
that the relation can be described, formally, as an \emph{adjunction}
from the category $PreOrd$ of preorders to the category $Cat$ of
small categories. The said adjunction mathematically expresses the
intuition that ``pre-orders approximate categories,'' asserting
thereby that \emph{every} concept in order theory has a corresponding
more general concept in category theory. For more on the relation
between order theory and category theory (\emph{i.e.}, metaphorically,
on $OT\rightarrow CT$) and on the correspondence between their concepts
see $\mathsection$\ref{sec:More-Abstract}.}
\end{itemize}
\end{itemize}

\subsubsection{References}

See~\cite{Pierce91,Rol2002,Geldrop2009,spivak2014category,Fong2018}.

\section{\label{sec:Comparison-Summary}Comparison Summary}

\begin{table*}
\noindent \centering{}%
\begin{tabular}{|c|c|c|c|}
\cline{2-4} 
\multicolumn{1}{c|}{} & \textbf{Order Theory} & \textbf{Set Theory} & \textbf{FP Type Theory}\tabularnewline
\hline 
Domain & Points of a Set $\mathbb{O}$ & $\mathbb{U}$=Subsets of a Set $U$ & Structural Data Types $\mathbb{D}$\tabularnewline
\hline 
Relation & Abstract Ordering $\leq$ & Inclusion $\subseteq$ & Inclusion $\subseteq$\tabularnewline
\hline 
Operator & Function $F:\mathbb{O}\rightarrow\mathbb{O}$ & Function $F:\mathbb{U}\rightarrow\mathbb{U}$ & Type Cons. $F:\mathbb{D}\rightarrow\mathbb{D}$\tabularnewline
\hline 
\multirow{2}{*}{Generator} & \multirow{2}{*}{$F$ Monotonic} & \multirow{2}{*}{$F$ Monotonic} & $F$ Poly. Type Constr.\tabularnewline
 &  &  & ($+,\times,\rightarrow$, and comp.)\tabularnewline
\hline 
\multirow{2}{*}{Pre-Fix Pt} & \emph{Pre-Fixed Point} $P$ & \emph{$F$-Large Set} $P$ & \emph{$F$-Closed Type} $P$\tabularnewline
 & $F\left(P\right)\leq P$ & $F\left(P\right)\subseteq P$ & $F\left(P\right)\subseteq P$\tabularnewline
\hline 
\multirow{2}{*}{Post-Fix Pt} & \emph{Post-Fixed Point} $P$ & \emph{$F$-Small Set} $P$ & \emph{$F$-Bounded Type} $P$\tabularnewline
 & $P\leq F\left(P\right)$ &  $P\subseteq F\left(P\right)$ & $P\subseteq F\left(P\right)$\tabularnewline
\hline 
 & If $\mathbb{O}$ Complete Lattice & \emph{Smallest $F$-Large Set} & \emph{Smallest $F$-Closed Type}\tabularnewline
Least FP & $\mu_{F}=\underset{F\left(P\right)\leq P}{\bigwedge}P$ & \multirow{2}{*}{ $\mu_{F}=\underset{F\left(P\right)\subseteq P}{\bigcap}P$} & $\mu_{F}=\underset{F\left(P\right)\subseteq P}{\bigcap}P$\tabularnewline
 & ($\bigwedge$ denotes `meet') &  & (Inductive Type)\tabularnewline
\hline 
 & If $\mathbb{O}$ Complete Lattice & \emph{Largest $F$-Small Set} & \emph{Largest $F$-Bounded Type}\tabularnewline
Grt. FP & $\nu_{F}=\underset{P\leq F\left(P\right)}{\bigvee}P$ & \multirow{2}{*}{$\nu_{F}=\underset{P\subseteq F\left(P\right)}{\bigcup}P$} & $\nu_{F}=\underset{P\subseteq F\left(P\right)}{\bigcup}P$\tabularnewline
 & ($\bigvee$ denotes `join') &  & (Coinductive Type)\tabularnewline
\hline 
Induction & \multirow{2}{*}{$F\left(P\right)\leq P$$\implies$$\mu_{F}\leq P$} & \multirow{2}{*}{$F\left(P\right)\subseteq P$$\implies$$\mu_{F}\subseteq P$} & \multirow{2}{*}{$F\left(P\right)\subseteq P$$\implies$$\mu_{F}\subseteq P$}\tabularnewline
Principle &  &  & \tabularnewline
\hline 
Coinduction & \multirow{2}{*}{$P\leq F\left(P\right)$$\implies$$P\leq\nu_{F}$} & \multirow{2}{*}{$P\subseteq F\left(P\right)$$\implies$$P\subseteq\nu_{F}$} & \multirow{2}{*}{$P\subseteq F\left(P\right)$$\implies$$P\subseteq\nu_{F}$}\tabularnewline
Principle &  &  & \tabularnewline
\hline 
Complete Latt. & \multirow{1}{*}{Sometimes} & \multirow{1}{*}{Always} & \multirow{1}{*}{Always}\tabularnewline
\hline 
Op. Generator & \multirow{1}{*}{Sometimes} & \multirow{1}{*}{Sometimes} & \multirow{1}{*}{Always}\tabularnewline
\hline 
\end{tabular}\protect\caption{\label{tab:Comparison-1}Comparison of concepts by subdiscipline (Part
1).}
\end{table*}
\begin{table*}
\noindent \begin{centering}
\begin{tabular}{|c|c|c|c|}
\cline{2-4} 
\multicolumn{1}{c|}{} & \textbf{OOP Type Theory} & \textbf{First-Order Logic} & \textbf{Category Theory}\tabularnewline
\hline 
Domain & Class Types $\mathbb{T}$ & Statements $\mathbb{S}$ & Objects $\mathbb{O}$\tabularnewline
\hline 
Relation & Subtyping $<:$ & Implication $\Rightarrow$ & Arrow $\rightarrow$\tabularnewline
\hline 
Operator & Gen. Class $F:\mathbb{T}\rightarrow\mathbb{T}$ & Operator $F:\mathbb{S}\rightarrow\mathbb{S}$ & Functor $F:\mathbb{O}\rightarrow\mathbb{O}$\tabularnewline
\hline 
\multirow{2}{*}{Generator} & \multirow{2}{*}{$F$ Covariant} & \multirow{2}{*}{$F$ Covariant} & \multirow{2}{*}{$F$ Covariant}\tabularnewline
 &  &  & \tabularnewline
\hline 
\multirow{2}{*}{Pre-Fix Pt} & \emph{$F$-Supertype $P$} & \emph{$F$-Weak $P$} & \emph{$F$-Algebra $P$}\tabularnewline
 & $F\left(P\right)<:P$ & $F\left(P\right)\Rightarrow P$ & $F\left(P\right)\rightarrow P$\tabularnewline
\hline 
\multirow{2}{*}{Post-Fix Pt} & \emph{$F$-Subtype $P$} & \emph{$F$-Strong $P$} & \emph{$F$-Coalgebra $P$}\tabularnewline
 & $P<:F\left(P\right)$ & $P\Rightarrow F\left(P\right)$ & $P\rightarrow F\left(P\right)$\tabularnewline
\hline 
 & \emph{Least $F$-Supertype} & \emph{Strongest $F$-Weak Stmt} & \emph{Initial $F$-Algebra}\tabularnewline
Least FP & $\mu_{F}\thickapprox\underset{F\left(P\right)<:P}{\bigwedge}P$ & \multirow{1}{*}{$\mu_{F}=\underset{F\left(P\right)\Rightarrow P}{\bigwedge}P$} & $\thickapprox\mu_{F}$\tabularnewline
 & ($\bigwedge$ denotes `meet') & ($\bigwedge$ denotes conjunction) & \tabularnewline
\hline 
 & \emph{Greatest $F$-Subtype} & \emph{Weakest $F$-Strong Stmt} & \emph{Final $F$-Coalgebra}\tabularnewline
Grt. FP & $\nu_{F}\thickapprox\underset{P<:F\left(P\right)}{\bigvee}P$ & \multirow{1}{*}{$\nu_{F}=\underset{P\Rightarrow F\left(P\right)}{\bigvee}P$} & $\thickapprox\nu_{F}$\tabularnewline
 & ($\bigvee$ denotes `join') & ($\bigvee$ denotes disjunction) & \tabularnewline
\hline 
Induction & \multirow{2}{*}{$F\left(P\right)<:P$$\implies$$\mu_{F}<:P$} & \multirow{2}{*}{$F\left(P\right)\Rightarrow P$$\implies$$\mu_{F}\Rightarrow P$} & \multirow{2}{*}{$F\left(P\right)\rightarrow P$$\implies$$\mu_{F}\rightarrow P$}\tabularnewline
Principle &  &  & \tabularnewline
\hline 
Coinduction & \multirow{2}{*}{$P<:F\left(P\right)$$\implies$$P<:\nu_{F}$} & \multirow{2}{*}{$P\Rightarrow F\left(P\right)$$\implies$$P\Rightarrow\nu_{F}$} & \multirow{2}{*}{$P\rightarrow F\left(P\right)$$\implies$$P\rightarrow\nu_{F}$}\tabularnewline
Principle &  &  & \tabularnewline
\hline 
Complete Latt. & \multirow{1}{*}{Rarely} & \multirow{1}{*}{Always} & \multirow{1}{*}{Sometimes}\tabularnewline
\hline 
Op. Generator  & \multirow{1}{*}{Sometimes} & \multirow{1}{*}{Sometimes} & \multirow{1}{*}{Sometimes}\tabularnewline
\hline 
\end{tabular}
\par\end{centering}

\protect\caption{\label{tab:Comparison-2}Comparison of concepts by subdiscipline (Part
2).}
\end{table*}
Table~\vref{tab:Comparison-1} and Table~\vref{tab:Comparison-2}
summarize the formulations of the induction/\-co\-induc\-tion principles
and concepts related to them that we presented in~$\mathsection$\ref{sec:Order-Theory}-$\mathsection$\ref{sec:Category-Theory}.

\section[A More Abstract Treatment]{\label{sec:More-Abstract}A Fundamental and More Abstract Treatment}

As we hinted to in $\mathsection$\ref{sec:Category-Theory}, order
theory and category theory are strongly related. In fact the connection
between the two fields goes much, much further than we hinted at.%
\begin{comment}
TODO: Revise. It is indeed the case that almost all concepts in order
theory have corresponding, usually more general, concepts in category
theory, and almost all concepts in category theory have corresponding
more specific concepts (or concept ``rudiments'') in order theory.
Based on that, it can be easily said that order theory approximates
category theory, and that category theory generalizes order theory
(\emph{i.e.}, somewhat in an imprecise sense we have ``$OT\leq CT$''),
just as, say, the real numbers generalize the integers.\footnote{Readers familiar with group theory may note that category theory also
generalizes group theory, groupoid theory, and monoid theory (every
group is simultaneously a groupoid and a monoid). As such, order theory
and group/monoid/groupoid theory can be viewed as rather orthogonal
instantiations of category theory (order theory allows multiple objects,
but only single arrows between objects at most. Monoid theory requires
and allows exactly one object, but allows multiple arrows. Groupoid
theory allows multiple objects and multiple arrows, but requires all
arrows to be invertible. Group theory requires and allows exactly
one object, and it allows multiple arrows but requires them to be
invertible). From a category-theoretic point of view, the main difference
(TODO:true?) between objects and arrows is that all arrows can be
composed/combined, but (without further knowledge) objects are ``inert''/uncomposable/uncombinable.}
\end{comment}

\paragraph{Closure and Kernel Operators}

In order theory a \emph{closure operator} over a poset $\mathbb{O}$
is an idempotent extensive generator~\cite{Davey2002}. An \emph{extensive}
(or \emph{inflationary}) endofunction over $\mathbb{O}$ is an endofunction
$cl$ where 
\[
\forall P\in\mathbb{O}.P\leq cl\left(P\right),
\]
meaning that all points of $\mathbb{O}$ are $cl$-small (\emph{i.e.},
are post-fixed points of $cl$). An endofunction $cl$\emph{ }is \emph{idempotent}
iff 
\[
\forall P\in\mathbb{O}.cl\left(cl\left(P\right)\right)=cl\left(P\right),
\]
\emph{i.e.}, iff $cl^{2}=cl$, meaning that applying $cl$ twice does
not transform or change an element of $\mathbb{O}$ any more than
applying $cl$ to the element once does.

Also, in order theory a \emph{kernel }(or \emph{interior}) \emph{operator}
over $\mathbb{O}$ is an idempotent intensive generator. An \emph{intensive
}(or \emph{deflationary}) endofunction over $\mathbb{O}$ is an endofunction
$kl$ where
\[
\forall P\in\mathbb{O}.kl\left(P\right)\leq P,
\]
meaning that all points of $\mathbb{O}$ are $kl$-large (\emph{i.e.},
are pre-fixed points of $kl$).\footnote{It may be helpful here to check Figure~1 of~\cite{AbdelGawad2019a}.
If $F$ in Figure~1 of~\cite{AbdelGawad2019a} is a closure operator
$cl$ then the upper diamond minus the inner diamond ``collapses''
(becomes empty), and we have $\nu_{cl}=\top=U$. That is because,
since $cl$ is extensive, we have $cl\left(\top\right)=\top$, and
thus $\top$---if it exists in $\mathbb{O}$---is always a \emph{fixed}
point of $cl$, in fact the \emph{gfp} (greatest fixed point) of $cl$.
Dually, if $F$ in Figure~1 of~\cite{AbdelGawad2019a} is a kernel
operator $kl$ then the lower diamond minus the inner diamond ``collapses'',
and we have $\mu_{kl}=\bot=\phi$. That is because, since $kl$ is
intensive, we have $kl\left(\bot\right)=\bot$, and thus $\bot$---if
it exists in $\mathbb{O}$---is always a fixed point of $kl$, in
fact the \emph{lfp} (least fixed point) of $kl$.}

\paragraph{Monads and Comonads}

In category theory, on the other hand, when a partially-ordered set
$\mathbb{O}$ is viewed as a category, then a \emph{monad} on a poset
$\mathbb{O}$ turns out to be exactly a closure operator over $\mathbb{O}$~\cite{spivak2014category}%
\begin{comment}
 (in logic a monad/closure operator is called a \emph{modal operator})
\end{comment}
. By the definition of monads, all objects of a category are coalgebras
of a monad, which translates to all elements of a poset $\mathbb{O}$
being post-fixed points of the corresponding closure operator on $\mathbb{O}$
(which we noted above). As such, the \emph{algebras} for this monad
correspond to the pre-fixed points of the closure operator, and thus
correspond exactly to its fixed points.

Similarly, a \emph{comonad} (the dual of a monad) on $\mathbb{O}$
turns out to be exactly a kernel operator over $\mathbb{O}$. As such,
by a dual argument, the \emph{coalgebras} for this comonad are the
post-fixed points of the kernel operator, and thus also correspond
exactly to its fixed points.

\begin{comment}
TODO: Revise, and put in Appendix? Further, an \emph{adjunction} in
category theory between two posets $\mathbb{O}$ and $\mathbb{Q}$
(viewed as categories) turns out to be exactly a Galois connection
between the two posets (category theory is where the term `adjoint'
of order theory comes from, and where the lower adjoint of Galois
connection becomes the \emph{left adjoint} of the corresponding adjunction
and the upper adjoint of the connection becomes the \emph{right adjoint}
of the adjunction).
\end{comment}

\paragraph{An Alternative Presentation}

Based on these observations that further relate category theory and
order theory, the whole technical discussion in this article can be
presented, more succinctly if also more abstractly, by rewriting it
in terms of the very general language of monads/comonads, then specializing
the abstract argument by applying it to each of the six categories
we are interested in.\footnote{Namely, (1) category \textbf{OrdE} of partially-ordered sets together
with self-maps, (2) category \textbf{PSetE} of power sets (ordered
by inclusion) together with endofunctions, (3) category \textbf{TypS}
of structural types (ordered by inclusion/structural subtyping) together
with structural type constructors, (4) category \textbf{TypN} of nominal
types (ordered by nominal subtyping) together with generic classes,
(5) category \textbf{FOL} of first-order logical statements (ordered
by implication) together with logical operators, and (6) category
\textbf{CatE} of (small) categories (ordered/related by arrows) together
with endofunctors.} Given that our goal, however, is to compare the concepts in their
most natural and most concrete mathematical contexts, we refrain from
presenting such a fundamental treatment here, keeping the possibility
of making such a presentation in some future work, \emph{e.g.}, as
a separate article, or as an %
\begin{comment}
currently underway 
\end{comment}
appendix to future versions of this article.

\subsubsection{References}

See~\cite{Rol2002,spivak2014category,Fong2018}.

\section*{Acknowledgments}

The author would like to thank John Greiner (Rice University) and
David Spivak (MIT) for their suggestions and their valuable feedback
on earlier versions of this article.

\bibliographystyle{plain}

\end{document}